# Exact relativistic treatment of stationary counter-rotating dust disks III Physical Properties

J. Frauendiener,

Max-Planck-Institut für Mathematik in den Naturwissenschaften,

Inselstr. 22–26,

04103 Leipzig, Germany,

C. Klein,

Laboratoire de Gravitation et Cosmologie Relativistes,

Université P. et M. Curie,

4, place Jussieu, 75005 Paris, France.

**Abstract**

This is the third in a series of papers on the construction of explicit solutions to the stationary axisymmetric Einstein equations which can be interpreted as counter-rotating disks of dust. We discuss the physical properties of a class of solutions to the Einstein equations for disks with constant angular velocity and constant relative density which was constructed in the first part. The metric for these spacetimes is given in terms of theta functions on a Riemann surface of genus 2. It is parameterized by two physical parameters, the central redshift and the relative density of the two counter-rotating streams in the disk. We discuss the dependence of the metric on these parameters using a combination of analytical and numerical methods. Interesting limiting cases are the Maclaurin disk in the Newtonian limit, the static limit which gives a solution of the Morgan and Morgan class and the limit of a disk without counter-rotation. We study the mass and the angular momentum of the spacetime. At the disk we discuss the energy-momentum tensor, i.e. the angular velocities of the dust streams and the energy density of the disk. The solutions have ergospheres in strongly relativistic situations. The ultrarelativistic limit of the solution in which the central redshift diverges is discussed in detail: In the case of two counter-rotating dust components in the disk, the solutions describe a disk with diverging central density but finite mass. In the case of a disk made up of one component, the exterior of the disks can be interpreted as the extreme Kerr solution.

# 1 Introduction

Relativistic dust disks have been studied since the late sixties [1], the reasons for the interest in these configurations being both physical and mathematical. The physical motivation arises from the importance of disk-shaped matter distributions in certain galaxies and accretion disks. Whereas general relativistic effects do not play a role in the context of galaxies, they have to be taken into account in the case of disks around black-holes since black-holes are genuinely relativistic objects. Moreover disks can be considered as limiting configurations of fluid bodies for vanishing pressure (see e.g. [2]). From a more mathematical point of view, dust disks offer the opportunity to obtain global spacetimes containing matter distributions which can be physically interpreted. The Einstein equations for an ideal fluid do not seem to be integrable even in the stationary axisymmetric case. Infinitesimally thin disks provide a possibility to circumvent this problem because the matter is reduced to two spatial dimensions. This leads to ordinary differential equations inside the disk which can be integrated at least in principle. Consequently one has to solve a boundary value problem for the vacuum equations where the boundary data follow from the properties of the matter in the disk. Since dust disks have no radial pressures one can place the disks without loss of generality in the equatorial plane even in the standard Weyl coordinates. Thus one avoids the complications of a free boundary value problem where the location of the disk has to be determined as part of the solution of the boundary value problem. The first solutions for relativistic dust disks were given by Morgan and Morgan [1]. They considered static spacetimes with disks which can be interpreted as being made up of two counter-rotating dust streams with vanishing total angular momentum. Bardeen and Wagoner [2] studied numerically a uniformly rotating disk consisting of a single dust component and as a post-Newtonian expansion. They compared this stationary solution to the Einstein equations to the static and the Newtonian case and gave a detailed discussion of the physical features of the spacetime. Later Neugebauer and Meinel [3] gave an explicit solution for the Bardeen-Wagoner disk in terms of Korotkin's solutions [4, 5] on a Riemann surface of genus 2 (in [6] it was shown that the solution [3] belongs to the class [4]).

In the first paper of this series [7] (henceforth referred to as I) we studied stationary counter-rotating dust disks and their relation to hyperelliptic functions. As an example of this approach we gave an explicit solution on a Riemann surface of genus 2 [8] where the two counter-rotating dust streams have constant angular velocity and constant relative density. In the limit of only one component one gets the solution of [3], in the limit of identical densities one gets a static solution of the Morgan and Morgan class. In the second paper [9] (henceforth referred to as II) we gave explicit formulas for the Ernst potential at the axis and the disk which are needed to discuss the energy-momentum tensor and considered limiting cases.

In the present paper we discuss the physical features of the hyperelliptic solutions [10, 11] which are a subclass of Korotkin's finite gap solutions [4, 5] in the example of the solution of I. We demonstrate how one can extract physically interesting quantities from the hyperelliptic functions in terms of which the metric is given. The solutions are explicit i.e. all metric functions are given in terms of quadratures and a set of well-defined functions, the theta functions. The integrals are evaluated numerically by making use of pseudospectral techniques. The metric depends on two physical parameters: $\varepsilon = z_R/(1+z_R)$ is related to the redshift $z_R$ of photons emitted from the center of the disk and detected at infinity; $\gamma$ is the relative density of the counter-rotating streams in the disk. In the Newtonian limit $\varepsilon$ is approximately 0 whereas it tends to 1 in the ultrarelativistic limit where the central redshift diverges. The limit of a single component disk is reached for $\gamma = 1$ (we will only consider positive values of $\gamma$), the static limit for $\gamma = 0$.

We give analytic expressions for the mass and the angular momentum as an expansion of the metric functions at infinity and as an integral over the energy-momentum tensor at the disk. The resulting analytic expressions have to be identical which provides a test for the numerics. In [12] Bičák and Ledvinka considered infinite disks of finite mass as sources for the Kerr metric. It was shown that the matter in the disk can be interpreted either as a disk with purely azimuthal stresses or as a disk with two counter-rotating dust components if the energy-conditions are satisfied. The same discussion is possible in the case considered here. As in [12] we discuss the matter in the disk using observers which rotate in a way that the energy-momentum tensor is diagonal for them. We study the angular velocity of these observers with respect to the locally non-rotating frames, and the angular velocities and the energy densities of the dust components which these observers measure. In the limit of diverging central redshift the spacetime is no longer asymptotically flat in the case of a one component disk, and the axis is no longer elementary flat. This behavior can be related as in [2] to the vanishing of the radius $\rho_0$ of the disk which was used as a length scale. If one carries out the limit $\rho_0 \to 0$ for $\rho \neq 0$, the metric becomes the extreme Kerr metric. In this limit the disk vanishes behind the horizon of the extreme Kerr solution. In the case of two counter-rotating dust components the radius of the disk remains finite even in the limit where the central redshift diverges. In the ultrarelativistic limit of the static disks, the matter in the disk moves at the speed of light, the energy density diverges at the center of the disk but the mass remains finite.

We closely follow the discussion in the pioneering paper [2], but this time for a class of solutions which depend on two parameters which continuously interpolate between the Newtonian and the ultrarelativistic regime, and the static and the Bardeen-Wagoner case respectively. The paper is organized is follows: In section 2 we summarize results of I and II and write down the complete metric corresponding to the Ernst potential of I in terms of theta functions. We outline the numerical scheme and present typical plots for the metric functions. In section 3 we discuss various physical properties of the solutions: We relate the physical parameters $\varepsilon$ and $\gamma$ to the parameters on which the analytic solution depends and discuss mass and angular momentum. The angular velocity $\Omega$ is discussed as a function of $\varepsilon$ and $\gamma$. We study the energy-momentum tensor at the disk as in [12] as well as the occurrence of ergospheres. In section 4 we discuss the ultrarelativistic limit of the solutions. We briefly discuss the over-extreme case for the one-component solution where the boundary value problem at the disk is still solved but where a ring singularity exists in the spacetime since the parameters of the solution are beyond the ultrarelativistic limit. In section 5 we add some concluding remarks.

# 2 Metric functions

## 2.1 Ernst potential and metric

We will briefly summarize results of I and II where details of the notation can be found. We use the Weyl–Lewis–Papapetrou metric (see e.g. [13])

$$ds^2 = -e^{2U}(dt + a\,d\phi)^2 + e^{-2U}\left(e^{2k}(d\rho^2 + d\zeta^2) + \rho^2 d\phi^2\right), \tag{2.1}$$

where $\rho$ and $\zeta$ are Weyl's canonical coordinates and $\partial_t$ and $\partial_\phi$ are the two commuting asymptotically timelike respectively spacelike Killing vectors. With $z = \rho + i\zeta$ and the potential $b$ defined by

$$b_z = -\frac{i}{\rho}e^{4U}a_z, \tag{2.2}$$

and $b \to 0$ for $z \to \infty$, we define the complex Ernst potential $f = e^{2U} + ib$ which is subject to the Ernst equation [14]

$$f_{z\bar{z}} + \frac{1}{2(z+\bar{z})}(f_{\bar{z}} + f_z) = \frac{2}{f+\bar{f}}f_z f_{\bar{z}}, \tag{2.3}$$

where a bar denotes complex conjugation in $\mathbb{C}$. The metric function $k$ follows from

$$k_z = 2\rho \frac{f_z \bar{f}_z}{(f+\bar{f})^2}. \tag{2.4}$$

In I (section 3) we have considered disks which can be interpreted as two counter-rotating components of pressureless matter, so-called dust. The surface energy-momentum tensor $S^{\mu\nu}$ of these models is defined on the hypersurface $\zeta = 0$. The tensor $S^{\mu\nu}$ is related to the energy-momentum tensor $T^{\mu\nu}$ which appears in the Einstein equations $G^{\mu\nu} = 8\pi T^{\mu\nu}$ (we use units in which the Newtonian gravitational constant and the velocity of light are equal to 1) via $T^{\mu\nu} = S^{\mu\nu}e^{k-U}\delta(\zeta)$. The tensor $S^{\mu\nu}$ can be written in the form

$$S^{\mu\nu} = \sigma_+ u_+^\mu u_+^\nu + \sigma_- u_-^\mu u_-^\nu, \tag{2.5}$$

where greek indices stand for the $t$, $\rho$ and $\phi$ components and where $u_\pm = (1, 0, \pm\Omega)$. A physical interpretation of this tensor will be given in section 3. We gave an explicit solution for disks with constant angular velocity $\Omega$ and constant relative density $\gamma = (\sigma_+ - \sigma_-)/(\sigma_+ + \sigma_-)$. This class of solutions is characterized by two real parameters $\lambda$ and $\delta$ which are related to $\Omega$ and $\gamma$ and the metric potential $U_0$ at the center of the disk via

$$\lambda = 2\Omega^2 e^{-2U_0} \tag{2.6}$$

and

$$\delta = \frac{1-\gamma^2}{\Omega^2}. \tag{2.7}$$

We put the radius $\rho_0$ of the disk equal to 1 unless otherwise noted. Since the radius appears only in the combinations $\rho/\rho_0$, $\zeta/\rho_0$ and $\Omega\rho_0$ in the physical quantities it does not have an independent role. It is always possible to use it as a natural length scale unless it tends to 0 as in the case of the ultrarelativistic limit of the one component disk. The Ernst potential will be discussed in dependence of the parameters $\varepsilon = z_R/(1+z_R) = 1 - e^{U_0}$ and $\gamma$.

The solution of the Ernst equation corresponding to the above energy-momentum tensor is given on a hyperelliptic Riemann surface $\Sigma_2$ of genus 2 which is defined by the algebraic relation $\mu^2(K) = (K+iz)(K-i\bar{z})\prod_{i=1}^{2}(K-E_i)(K-\bar{E}_i)$ (see I, section 4 for details of the notation). We choose $\operatorname{Re} E_1 < 0$, $\operatorname{Im} E_i < 0$ and $E_1 = -\bar{E}_2$ with $\bar{E}_2 = \alpha_1 + i\beta_1$. We use the cut-system of Fig. 1 for the numerical calculations since it is adapted to the symmetry of the problem. The base point of the Abel map is $E_1$.

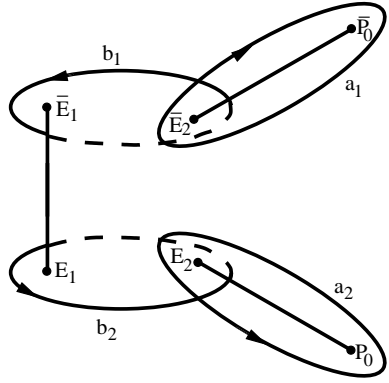

Figure 1: Cut-system.

In this cut-system the solution of I (Theorem 7.2) takes the form

$$f(\rho, \zeta) = \frac{\Theta[m](\omega(\infty^+) + u)}{\Theta[m](\omega(\infty^+) - u)} e^I, \qquad (2.8)$$

where $\Theta[m]$ is the theta function on $\Sigma_2$ with half-integer characteristic $[m]$, where $I = \frac{1}{2\pi i}\int_\Gamma \ln G(\tau) d\omega_{\infty^+\infty^-}(\tau)$, where $u_i = \frac{1}{2\pi i}\int_\Gamma \ln G d\omega_i$, where $\Gamma$ is the covering of the imaginary axis in the +-sheet of $\Sigma_2$ between $-i$ and $i$, where the characteristic $[m] = \begin{bmatrix} 1 & 0 \\ 1 & 0 \end{bmatrix}$, and where

$$G(\tau) = \frac{\sqrt{(\tau^2 - \alpha)^2 + \beta^2} + \tau^2 + 1}{\sqrt{(\tau^2 - \alpha)^2 + \beta^2} - (\tau^2 + 1)}. \qquad (2.9)$$

The branch points of the Riemann surface are given by the relation $E := E_1^2 = \alpha + i\beta$ with $\alpha$, $\beta$ real and

$$\alpha = -1 + \frac{\delta}{2}, \quad \beta = \sqrt{\frac{1}{\lambda^2} + \delta - \frac{\delta^2}{4}}. \qquad (2.10)$$

Regularity of the solutions restricts the range of the physical parameters to $0 \leq \delta \leq \delta_s(\lambda) := 2\left(1 + \sqrt{1 + 1/\lambda^2}\right)$ and $0 < \lambda \leq \lambda_c$ where $\lambda_c(\gamma)$ is the smallest value of $\lambda$ for which $\varepsilon = 1$. We note that with $\alpha$ and $\beta$ given, the Riemann surface is completely determined at a given point in the spacetime, i.e. for a given value of $P_0$. The dependence of the solution (2.8) on the physical coordinates is exclusively through the branch points $P_0$ and $\bar{P}_0$.

The complete metric (2.1) can be expressed via theta functions (see II, Theorem 2.2 where a different cut-system is used). With the characteristics $[n_i]$ given by

$$[n_1] = \begin{bmatrix} 1 & 1 \\ 1 & 1 \end{bmatrix}, \quad [n_2] = \begin{bmatrix} 0 & 0 \\ 1 & 1 \end{bmatrix}, \quad [n_3] = \begin{bmatrix} 1 & 0 \\ 1 & 0 \end{bmatrix}, \quad [n_4] = \begin{bmatrix} 0 & 1 \\ 1 & 0 \end{bmatrix}, \quad (2.11)$$

the function $e^{2U}$ can be written in the form

$$e^{2U} = \frac{\Theta[n_1](u)\Theta[n_2](u)}{\Theta[n_1](0)\Theta[n_2](0)} \frac{\Theta[n_3](\omega(\infty^-))\Theta[n_4](\omega(\infty^-))}{\Theta[n_3](\omega(\infty^-)+u)\Theta[n_4](\omega(\infty^-)+u)} e^I. \quad (2.12)$$

The function $e^{2U}$ which is just the real part of the Ernst potential was written in [11] in the form (2.12) with the help of Fay's trisecant identity [15]. This form is especially adapted for determining ergospheres which are just the zeros of $e^{2U}$. In [11] it was shown that the real part of the Ernst potential can only vanish if $\Theta[n_1](u)\Theta[n_2](u) = 0$ which provides a necessary condition for the occurrence of ergospheres (the sufficient condition is that the denominator in (2.12) is non-zero in this case).

Korotkin [4] gave an expression for the metric function $a$ as a derivative of theta functions with respect to the argument. In [11] this formula could be written in the form (2.13) free off derivatives by using the trisecant identity which leads to

$$(a-a_0)e^{2U} = -\rho \left( \frac{\Theta[n_1](0)\Theta[n_2](0)}{\Theta[n_3](\omega(\infty^-))\Theta[n_4](\omega(\infty^-))} \frac{\Theta[n_1](u)\Theta[n_2](u+2\omega(\infty^-))}{\Theta[n_3](u+\omega(\infty^-))\Theta[n_4](u+\omega(\infty^-))} - 1 \right), \quad (2.13)$$

where the constant $a_0 = -\gamma/\Omega$. The constant can be expressed via theta functions on the elliptic surface $\Sigma'$ given by $\mu'^2 = (K-E_1^2)(K-\bar{E}_1^2)$ (see [11], II). We denote quantities defined on $\Sigma'$ by a prime and get

$$a_0 = \frac{\beta_1}{\alpha_1}\sqrt{\alpha_1^2+\beta_1^2} \left( \frac{\vartheta_4^2(0)}{\vartheta_3(\omega'(\infty^-))\vartheta_4(\omega'(\infty^-))} \right)^2 \frac{\vartheta_4(u'+2\omega'(\infty^-))}{\vartheta_4(u')} e^{-I'}, \quad (2.14)$$

where $d\omega_1 = d\omega'$, $d\omega_2 = d\omega'_{\zeta^-\zeta^+}$, $u_i = \frac{1}{2\pi i}\int_\Gamma \ln G d\omega_i$, and where $I' = \frac{1}{2\pi i}\int_\Gamma \ln G d\omega'_{\infty^+\infty^-}$. The elliptic theta functions $\vartheta_i$ where $i = 1,\ldots,4$ have the characteristics $\begin{bmatrix} 1 \\ 1 \end{bmatrix}$, $\begin{bmatrix} 1 \\ 0 \end{bmatrix}$, $\begin{bmatrix} 0 \\ 0 \end{bmatrix}$ and $\begin{bmatrix} 0 \\ 1 \end{bmatrix}$ respectively.

Whereas the metric functions $a$ and $e^{2U}$ can be invariantly expressed through the scalar products of the Killing vectors, this is not the case for the metric function $e^{2k}$. Nonetheless it is interesting to know this function because it determines the geometry of the $(\rho,\zeta)$-space and because of its relation to the $\tau$-function of the linear system associated with the Ernst equation (see [16]). This connection made it possible to derive an explicit expression for $k$ in terms of theta functions of

(2.15) in [17]:

$$e^{2k} = C\frac{\Theta[n_1](u)\Theta[n_2](u)}{\Theta[n_1](0)\Theta[n_2](0)}\exp\left(\frac{2}{(4\pi i)^2}\int_\Gamma\int_\Gamma dK_1 dK_2 h(K_1)h(K_2)\ln\frac{\Theta_o(\omega(K_1)-\omega(K_2))}{K_1-K_2}\right),$$
(2.15)

where $\Theta_o$ is a theta function with an odd characteristic, where $h(\tau) = \partial_\tau \ln G(\tau)$, and where $C$ is a constant which is determined by the condition that $k$ vanishes on the regular part of the axis and at infinity. It reads

$$1/C = \frac{\vartheta_4^2(u')}{\vartheta_4^2(0)}\exp\left(\frac{2}{(4\pi i)^2}\int_\Gamma\int_\Gamma dK_1 dK_2 h(K_1)h(K_2)\ln\frac{\vartheta_1(\omega'(K_1)-\omega'(K_2))}{K_1-K_2}\right). \quad (2.16)$$

In an ergoregion, the function $\Theta[n_1](u)\Theta[n_2](u)$ becomes negative. Since the remaining terms in (2.15) cannot change sign, the function $e^{2k}$ is always negative where $e^{2U}$ is negative. The metric function $g_{11} = g_{22} = e^{2(k-U)}$ is consequently non-negative.

Since we can concentrate on positive values of $\zeta$ because of the equatorial symmetry of the solution, the Riemann surface can only become singular if $P_0$ coincides with $\bar{P}_0$, i.e. on the axis, or if it coincides with $E_2$. Coinciding branch points imply that some of the periods diverge. Although the Ernst potential is regular at the axis, this causes problems for the numerical evaluation which affect the accuracy. Therefore we substitute the analytic expression (see II, Theorem 3.1)

$$f(0,\zeta) = \frac{\vartheta_4(\int_{\zeta^+}^{\infty^+} d\omega' + u') - \exp(-\omega_2(\infty^+) - u_2)\vartheta_4(\int_{\zeta^-}^{\infty^+} d\omega' + u')}{\vartheta_4(\int_{\zeta^+}^{\infty^+} d\omega' - u') - \exp(-\omega_2(\infty^+) + u_2)\vartheta_4(\int_{\zeta^-}^{\infty^+} d\omega' - u')}e^{I'+u_2}. \quad (2.17)$$

The real part of the Ernst potential can be written in the form

$$e^{2U} = \frac{\vartheta_4^2(u')}{\vartheta_4^2(0)}\frac{\vartheta_4^2\left(\int_{\zeta^+}^{\infty^-} d\omega'\right) - \exp(-2\omega_2(\infty^-))\vartheta_4^2\left(\int_{\zeta^-}^{\infty^-} d\omega'\right)}{\vartheta_4^2\left(u' + \int_{\zeta^+}^{\infty^-} d\omega'\right) - \exp(-2\omega_2(\infty^-) - 2u_2)\vartheta_4^2\left(u' + \int_{\zeta^-}^{\infty^-} d\omega'\right)}. \quad (2.18)$$

With these analytic formulas on the axis, one can obtain accurate numerical results since, for $\zeta \neq 0$, the metric functions have an expansion of the form $F(\rho,\zeta) = F(0,\zeta) + \rho^2 F_2(\zeta) + 0(\rho^4)$ in the vicinity of the axis.

If $P_0$ coincides with $E_2$, the Ernst potential and the metric functions can be expressed in terms of quantities defined on the Riemann surface $\Sigma''$ of genus 0 given by $\mu''^2(\tau) = (\tau - E_1)(\tau - \bar{E}_1)$ i.e. via elementary functions (see II, Theorem 3.2). For $P_0 = E_2$ the differentials on $\Sigma_2$ reduce to differentials on $\Sigma''$, $d\omega_1 = d\omega''_{E_2^- E_2^+}$, $d\omega_2 = d\omega''_{\bar{E}_2^- \bar{E}_2^+}$ and $I = I'' = \frac{1}{2\pi i}\int_\Gamma \ln G d\omega''_{\infty^+\infty^-}$ where a double prime denotes that the quantity is defined on $\Sigma''$. The Ernst potential reads

$$f = \frac{\sinh\frac{\omega_1(\infty^+)+u_1}{2}}{\sinh\frac{\omega_1(\infty^+)-u_1}{2}}e^{I''}, \quad (2.19)$$

the function $a$ follows from

$$
\begin{aligned}
(a-a_0)e^{2U} &= \rho \left( \frac{\sinh \frac{\pi_{12}}{4}}{\sinh \frac{\omega_1(\infty^+)}{2} \sinh \frac{\omega_2(\infty^+)}{2}} \times \right. \\
&\left. \frac{\exp\left(\frac{\pi_{12}}{4}\right) \cosh \frac{u_1+u_2+2\omega_1(\infty^+)+2\omega_2(\infty^+)}{2} - \exp\left(-\frac{\pi_{12}}{4}\right) \cosh \frac{u_1-u_2+2\omega_1(\infty^+)-2\omega_2(\infty^+)}{2}}{2\sinh \frac{u_1-\omega_1(\infty^+)}{2} \sinh \frac{u_2-\omega_2(\infty^+)}{2}} - 1 \right),
\end{aligned}
\tag{2.20}
$$

and the function $e^{2k}$ is given by

$$
\begin{aligned}
e^{2k} &= C \frac{\exp\left(\frac{\pi_{12}}{4}\right) \cosh \frac{u_1+u_2}{2} - \exp\left(-\frac{\pi_{12}}{4}\right) \cosh \frac{u_1-u_2}{2}}{2\sinh \frac{\pi_{12}}{4}} \\
&\quad \exp\left( \frac{1}{(4\pi i)^2} \int_\Gamma \int_\Gamma \frac{dK_1 dK_2}{(K_1-K_2)^2} \ln G(K_1) \ln G(K_2) \times \right. \\
&\quad \left. \left( \sqrt{\frac{(K_1-E_1)(K_2-\bar{E}_1)}{(K_1-\bar{E}_1)(K_2-E_1)}} + \sqrt{\frac{(K_1-\bar{E}_1)(K_2-E_1)}{(K_1-E_1)(K_2-\bar{E}_1)}} - 2 \right) \right),
\end{aligned}
\tag{2.21}
$$

where $\pi_{12}$ is a component of the $b$-matrix on $\Sigma_2$.

At the disk the branch points $P_0, \bar{P}_0$ lie on the contour $\Gamma$ which implies that care has to be taken in the evaluation of the path integrals. The situation is however simplified by the equatorial symmetry of the solution which is reflected by the additional involution $K \to -K$ of the Riemann surface $\Sigma_2$ for $\zeta = 0$. This makes it possible to express the metric functions in terms of elliptic theta functions (see [11]). In II (Theorem 4.1) we could give especially efficient formulas for the functions needed to calculate the energy-momentum tensor at the disk. We denote with $\Sigma_w$ the elliptic Riemann surface defined by $\mu_w^2 = (\tau + \rho^2)((\tau - \alpha)^2 + \beta^2)$, and let $dw$ be the associated differential of the first kind with $u_w = \frac{1}{i\pi} \int_{-\rho^2}^{-1} \ln G(\sqrt{\tau}) dw(\tau)$. We cut the surface in a way that the $a$-cut is a closed contour in the upper sheet around the cut $[-\rho^2, \bar{E}]$ and that the $b$-cut starts at the cut $[\infty, E]$. The Abel map $w$ is defined for $P \in \Sigma_w$ as $w(P) = \int_\infty^P dw$. Then the real part of the Ernst potential at the disk can be written as

$$
\begin{aligned}
e^{2U} &= \frac{1}{Y-\delta} \left( -\frac{1}{\lambda} - \frac{Y}{\delta} \left( \frac{\frac{1}{\lambda^2}+\delta}{\sqrt{\frac{1}{\lambda^2}+\delta\rho^2}} - \frac{1}{\lambda} \right) \right. \\
&\quad \left. + \sqrt{\frac{Y^2((\rho^2+\alpha)^2+\beta^2)}{\frac{1}{\lambda^2}+\delta\rho^2} - 2Y(\rho^2+\alpha) + \frac{1}{\lambda^2}+\delta\rho^2} \right),
\end{aligned}
\tag{2.22}
$$

where

$$
Y = \frac{\frac{1}{\lambda^2}+\delta\rho^2}{\sqrt{(\rho^2+\alpha)^2+\beta^2}} \frac{\vartheta_3^2(u_w)}{\vartheta_1^2(u_w)}.
\tag{2.23}
$$

In I it was shown that there exist algebraic relations between the real and imaginary parts of the Ernst potential,

$$\frac{\delta^2}{2}(e^{4U}+b^2) = \left(\frac{1}{\lambda}-\delta e^{2U}\right)\left(\frac{\frac{1}{\lambda^2}+\delta}{\sqrt{\frac{1}{\lambda^2}+\delta\rho^2}}-\frac{1}{\lambda}\right)+\delta\left(\frac{\delta+\rho^2}{2}-1\right), \quad (2.24)$$

and the function $Z := (a-a_0)e^{2U}$

$$Z^2 - \rho^2 + \delta e^{4U} = \frac{2}{\lambda}e^{2U}. \quad (2.25)$$

At the rim of the disk ($\rho = 1$ and $\zeta = 0$) the value of the metric function $e^{2U}$ thus has the form

$$e^{2U(1,0)} = 1 - \frac{1}{\delta}\left(\sqrt{\frac{1}{\lambda^2}+\delta}-\frac{1}{\lambda}\right). \quad (2.26)$$

The imaginary part of the Ernst potential vanishes for $\gamma \neq 0$ at the rim of the disk as $(1-\rho^2)^{\frac{3}{2}}$. These explicit relations at the rim of the disk can be used as a test for the numerics.

## 2.2 Numerical evaluation of the hyperelliptic integrals

For the numerical evaluation of the above expressions we use pseudospectral methods. First the $a$- and $b$-periods of the hyperelliptic Riemann surface for the cut-system in Fig. 1 have to be determined. These are integrals between branch points $P_i$, $P_j$, $i \neq j$ of the Riemann surface,

$$\int_{P_i}^{P_j} \frac{\tau^n d\tau}{\mu(\tau)}, \quad n = 0, 1, 2. \quad (2.27)$$

With a linear transformation of the form $\tau = at + b$ they can be put into the form

$$\int_{-1}^{1} \frac{\alpha_0 + \alpha_1 t + \alpha_2 t^2}{\sqrt{1-t^2}} H(t)\, dt, \quad (2.28)$$

where the $\alpha_i$ are complex constants and where $H(t)$ is a continuous (in fact, analytic) complex valued function on the interval $[-1, 1]$. This form of the integral suggests to express the powers $t^n$ in terms of the first three Chebyshev polynomials $T_0(t) = 1$, $T_1(t) = t$ and $T_2(t) = 2t^2 - 1$ and to approximate the function $H(t)$ by a linear combination of Chebyshev polynomials

$$H(t) = \sum_{n \geq 0} h_n T_n(t).$$

Since the $T_n$ form a complete orthogonal system on the interval, this approximation can be made arbitrarily precise by using enough terms. Using the orthogonality relation between the Chebyshev polynomials

$$\int_{-1}^{1} T_n(t) T_m(t) \frac{dt}{\sqrt{1-t^2}} = \begin{cases} \pi & m = n = 0 \\ \pi/2 & m = n \neq 0 \\ 0 & m \neq n \end{cases} \quad (2.29)$$

the value of the integral is a linear combination of the coefficients $h_0$, $h_1$ and $h_2$. To determine these we have implemented a Fast Cosine Transform (FCT) within Matlab. It turns out that we can get accuracies of the order of the machine precision ($\approx 10^{-14}$) if we use 32, at most 128 terms in the approximating sum.

Since the sum of the *a*-periods and the integral over a closed contour around the cut $[E_1, \bar{E}_1]$ must exactly vanish, this can be used to test the numerics. When two or more branch points coincide as on the axis, the analytic expressions (2.17) to (2.21) are substituted.

The differentials $d\omega_i$ of the first kind are normalized by the condition $\oint_{a_j} d\omega_i = 2\pi i \delta_{ij}$, the differential $d\omega_{\infty^+\infty^-}$ of the third kind is normalized by the conditions that it has residues $+1$ and $-1$ at $\infty^+$ and $\infty^-$ respectively, and vanishing *a*-periods. The theta function is approximated by the sum

$$\Theta(x) = \sum_{n_1=-N}^{N} \sum_{n_2=-N}^{N} \exp\left(\frac{1}{2}\pi_{11}n_1^2 + \pi_{12}n_1n_2 + \frac{1}{2}\pi_{22}n_2^2 + n_1x_1 + n_2x_2\right). \tag{2.30}$$

The rapid convergence of the series due to negatively definite real part of $\Pi = \begin{pmatrix} \pi_{11} & \pi_{12} \\ \pi_{21} & \pi_{22} \end{pmatrix}$ makes it possible in general to obtain an accuracy of machine precision with values $N \leq 5$. To calculate the integrals $\omega(\infty^+)$ we use the fact (see e.g. [15]) that the *b*-periods of Abelian integrals of the third kind can be expressed via integrals of the first kind,

$$\oint_{b_i} d\omega_{\infty^+\infty^-} = \omega_i(\infty^+) - \omega_i(\infty^-). \tag{2.31}$$

These integrals are thus determined along with the *b*-periods of the integrals of the first kind. At the disk we use formulas (2.22) to (2.25). The non-Abelian integrals $u_i$, $I$ are determined also using pseudospectral methods. They can be written in the form

$$\int_{-1}^{1} dt H(t), \tag{2.32}$$

where $H(t)$ is a continuous complex-valued function on the interval $[-1, 1]$. The integration is performed by first approximating the integrand by a linear combination of Chebyshev polynomials as before. Then, making use of the identity

$$\frac{T'_{m+1}}{m+1} - \frac{T'_{m-1}}{m-1} = 2T_m \tag{2.33}$$

one can compute the expansion coefficients of a function $g$ on $[-1, 1]$ with $g' = H$ by applying the relation $2kg_k = h_{k-1} - h_{k+1}$ ($k > 0$) between the expansion coefficients. Finally, having transformed back, the value of the integral is obtained as $g(1) - g(-1)$.

In contrast to the algebro-geometric solutions of integrable equations like Korteweg-de Vries and Sine-Gordon (see e.g. [19]), the characteristic quantities of the Riemann surface as the periods have to be calculated at each point of the spacetime since the Ernst potential depends on the moving branch points $P_0$ and $\bar{P}_0$. Thus for each value of $(\rho, \zeta)$ one has to calculate nine integrals and to do the summation of the theta series to obtain the Ernst potential (2.8). Because of the equatorial symmetry, the calculation can be limited to $\zeta \geq 0$: whereas the metric functions are even in $\zeta$, the imaginary part of the Ernst potential is an odd function.

To illustrate the metric functions we show plots for $\varepsilon = 0.85$ and $\gamma = 0.99$ ($\lambda = 10.12$ and $\delta = 0.856$), i.e. a disk in a strongly relativistic situation. The metric function $e^{2U}$ (see Fig. 2) tends to 1 for large distances from the disk. At the disk it is continuous but its normal derivatives have a jump. In the vicinity of the disk, the function is negative which indicates the presence of an ergosphere. In the exterior of the disk, $e^{2U}$ is completely smooth and does not take a local extremum in the whole physical range of the parameters. The function thus shows the same analytic properties as a solution to the Laplace equation.

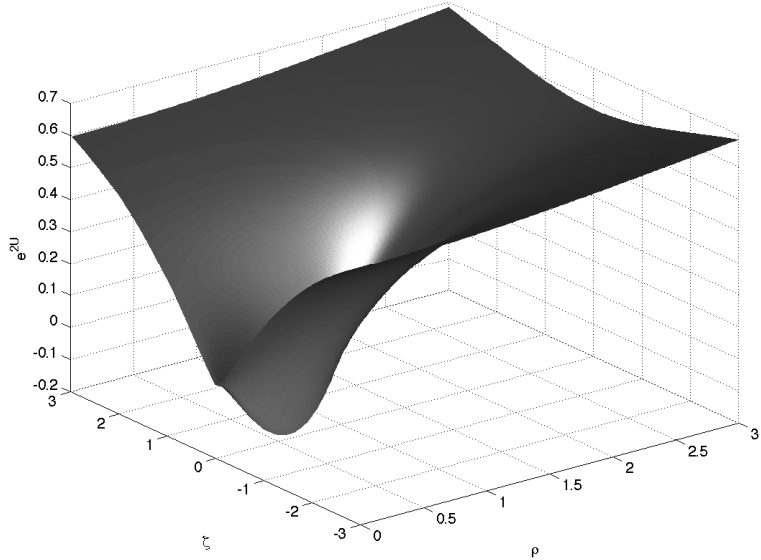

Figure 2: Metric function $e^{2U}$.

The imaginary part of the Ernst potential (see Fig. 3) is an odd function in $\zeta$. Thus it vanishes in the equatorial plane in the exterior of the disk. For large distances from the disk it tends to zero because of the asymptotic flatness of the spacetime. At the disk, the function has a jump which is zero at the rim of the disk since $b$ is continuous there.

The metric function $a$ (see Fig. 4) is equatorially symmetric and everywhere continuous. At the disk, the normal derivatives of $a$ have a jump, in the remaining spacetime it is completely regular. On the axis and at infinity the function is identically zero.

The function $e^{2k}$ in Fig. 5 has similar properties: it is equatorially symmetric and everywhere continuous, the normal derivatives have a jump at the disk. The function is identical to 1 on the axis ('elementary flatness') and at infinity (asymptotic flatness). The function is only significantly different from 1 in the vicinity of the disk. The metric function $e^{2(k-U)}$ is always positive even in the ergoregions which implies that the signature of the metric does not change.

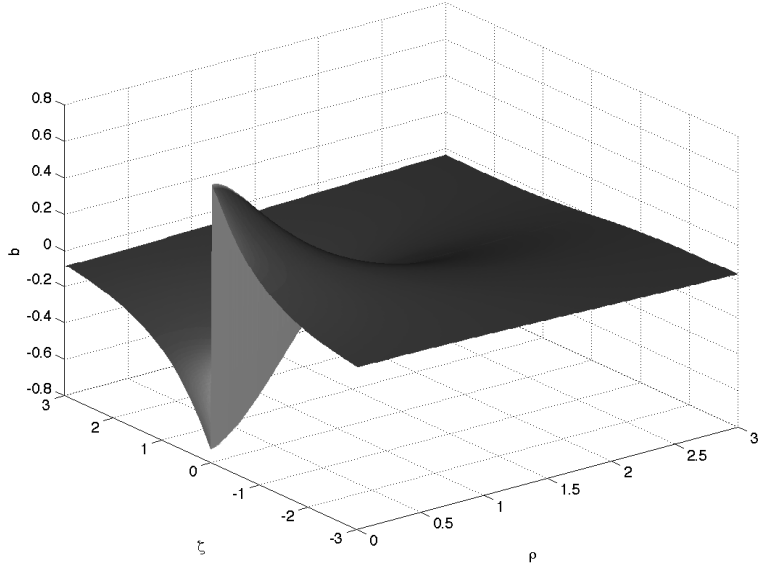

Figure 3: Imaginary part of the Ernst potential.

# 3 Physical properties

## 3.1 The physical parameters

We consider the metric as depending on the two physical parameters $\varepsilon$ and $\gamma$. Mathematically more natural are the parameters $\lambda$ and $\delta$. These two sets can be converted through the following procedure. The formula (2.18) can be used to calculate the real part of the Ernst potential at the origin, $e^{2U_0}$, which is related to the redshift $z_R$ of photons emitted from the center of the disk and detected at infinity, $z_R = e^{-U_0} - 1$,

$$e^{2U_0} = \frac{(1+X^2)(\sqrt{1+\lambda^2}-\lambda)}{X^2-(\sqrt{1+\lambda^2}-\lambda)^2}, \tag{3.1}$$

where $X$ is the purely imaginary quantity

$$X = \frac{\vartheta_3(u')\vartheta_4(0)}{\vartheta_1(u')\vartheta_2(0)}. \tag{3.2}$$

The corresponding values of $\lambda$ and $\delta$ follow from (2.6), (2.7) and (3.1). We get for $\varepsilon \neq 1$

$$\delta = \frac{1-\gamma^2}{(1-\varepsilon)^2}\frac{2}{\lambda}. \tag{3.3}$$

With this value we enter equation (3.1) for $e^{2U_0}$ and solve numerically for $\lambda(\varepsilon,\gamma)$. For $\delta = 0$ one finds that the first zero of $e^{2U_0}$ is reached for $\lambda_c(0) = 4.62966\ldots$. The function has additional zeros for higher values of $\lambda$ (see e.g. [10]). We are only interested in values $0 < \lambda < \lambda_c(\delta)$.

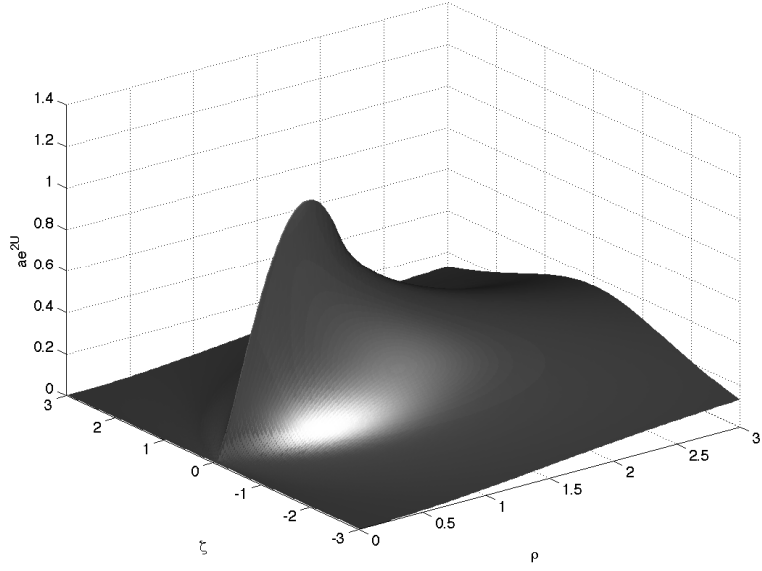

Figure 4: Metric function $ae^{2U}$.

For $\gamma < 1$ the quantity $e^{2U_0}$ is a monotonous function in $\lambda$ for $0 < \lambda < \infty$. Equation (3.3) then provides the corresponding value of $\delta(\varepsilon, \gamma)$.

For $\varepsilon = 1$ there are two cases: if $\gamma = 1$, then $\delta = 0$ and $\lambda = \lambda_c(0)$. For $\gamma \neq 1$, relation (3.3) implies that $\lambda_c(\delta)$ must be infinite. The corresponding value of $\delta$ follows with (2.6), (2.7) and (3.1) in the limit $\lambda \to \infty$ as the solution of the equation

$$\delta = \frac{4(1-\gamma^2)X^2}{1+X^2}. \tag{3.4}$$

Throughout the article we will consider the following limiting cases:

**Newtonian limit:** $\varepsilon = 0$ ($\lambda = 0$), i.e. small velocities $\Omega \rho_0$ and small redshifts in the disk. For $\lambda \to 0$, the integral $u'$ goes to zero. Thus the quantity $X$ diverges since $\vartheta_1$ is an odd function. Consequently one gets from (3.1) $U_0 = -\Omega^2$, the value for the Maclaurin disk (see II, Theorem 5.1). There it was shown that in this limit $e^{2U}$ tends to the Maclaurin disk solution, independently of $\gamma$. This solution can be written as

$$U(\rho, \zeta) = -\frac{1}{4\pi i} \int_{-i}^{i} \frac{2\lambda(\tau^2 + 1)}{\sqrt{(\tau - \zeta)^2 + \rho^2}} d\tau. \tag{3.5}$$

**ultrarelativistic limit:** $\varepsilon = 1$, i.e. diverging central redshift. For $\gamma = 1$ we have $\vartheta_4(u') = 0$ and thus $X = -i$ and $f_0 = -i$, i.e. the value of the Ernst potential of the extreme Kerr metric at the horizon. For $\gamma \neq 1$, the ultrarelativistic limit is reached for $\lambda \to \infty$.

**static limit:** $\gamma = 0$ ($\delta = \delta_s(\lambda)$). In this limit, the branch points of $\Sigma'$ collapse pairwise which leads to a diverging $X$ and $e^{2U_0} = \sqrt{1+\lambda^2} - \lambda$. In II (Theorem 5.2) it was shown that this is the Morgan and Morgan solution [1] for constant $\Omega$,

$$U(\rho, \zeta) = -\frac{1}{4\pi i} \int_{-i}^{i} \frac{\ln G(\tau)}{\sqrt{(\tau - \zeta)^2 + \rho^2}} d\tau \tag{3.6}$$

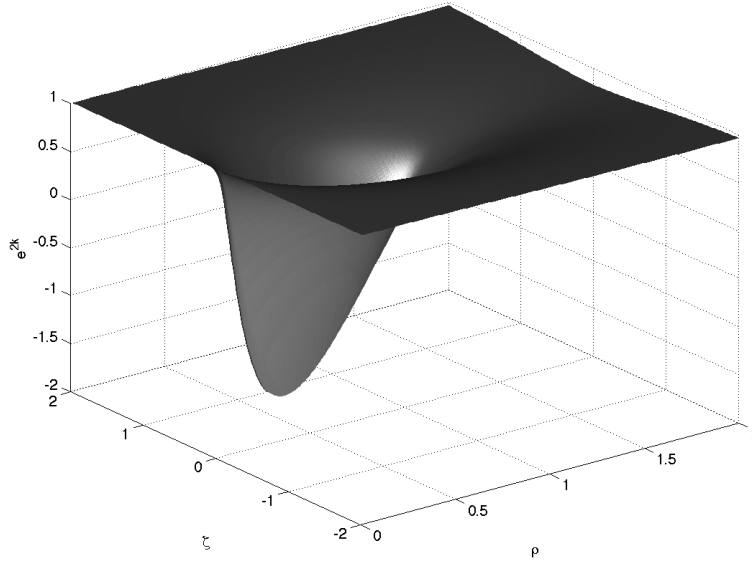

Figure 5: Metric function $e^{2k}$.

with
$$G = 1 - \frac{4}{\delta}(\tau^2 + 1). \tag{3.7}$$

At the disk one has
$$e^{2U} = \sqrt{\frac{1}{4} - \frac{1}{\delta}} + \sqrt{\frac{1}{4} - \frac{1}{\delta} + \frac{\rho^2}{\delta}}, \tag{3.8}$$

with $\Omega^2 \delta = 1$.

**one component:** $\gamma = 1$ ($\delta = 0$), i.e. no counter-rotating matter in the disk. This is the disk which was studied numerically by Bardeen and Wagoner [2]. The analytic solution is the solution by Neugebauer and Meinel [3] in the notation of [10].

The parameter $\lambda$ can be viewed as a 'relativity' parameter: for small values of $\lambda$, one is in the Newtonian regime, for larger values relativistic effects become more and more dominant up to the ultrarelativistic limit where the central redshift diverges. The values of $\lambda$ itself, however, do not have an invariant meaning. Thus it seems better to use the central redshift $z_R$ in $\varepsilon = z_R/(1 + z_R)$ as a parameter as in [2],
$$\varepsilon = 1 - e^{U_0}, \tag{3.9}$$
where $e^{U_0}$ is taken from (3.2).

In the ultrarelativistic limit, the values of $\delta$ must be between 0 (the one-component case) and 4 (the static limit, where $\gamma = 0$ and $X^2 \to \infty$). We plot $\varepsilon$ as a function of $\lambda$ for $\gamma = 1$ and $\gamma = 0$ in Fig. 6. In the case $\gamma = 1$, the function goes to 1 at finite values of $\lambda$ whereas for $\gamma \neq 1$ it goes monotonically to 1 as $\lambda$ goes to infinity as in the static case $\gamma = 0$.

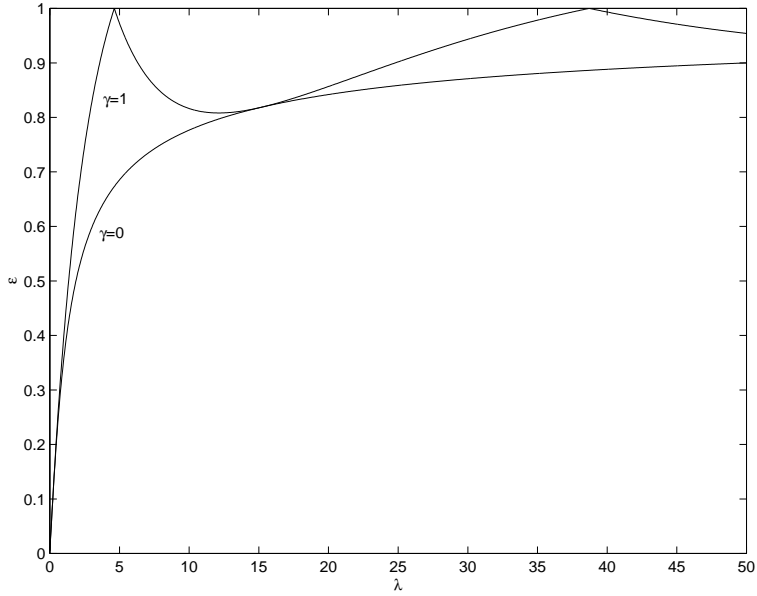

Figure 6: The function $\varepsilon$ in dependence of $\lambda$ for $\gamma = 1$ and $\gamma = 0$.

## 3.2 Mass and angular momentum

The ADM mass $M$ and the angular momentum $J$ of the spacetime (see e.g. [18]) can be obtained by expanding the axis potential (2.17) in the vicinity of infinity. The real part of the Ernst potential for $\varepsilon < 1$ reads $e^{2U} = 1 - 2M/\zeta + o(1/\zeta)$ and the imaginary part $b = 2J/\zeta^2 + o(1/\zeta^2)$. In II (Corollary 3.2) it was shown that the ADM mass is given by the formula

$$M = -D_{\infty^-} \ln \vartheta_4(u') - \frac{1}{4\pi i} \int_\Gamma \ln G \, d\omega_{1,\infty^+}, \tag{3.10}$$

and that the angular momentum is given by

$$J = -\frac{\gamma}{\Omega}\left(D_{\infty^-} \ln \vartheta_4(u') + D_{\infty^-} \ln \vartheta_2(u') + \frac{1}{2\pi i}\int_\Gamma \ln G \, d\omega_{1,\infty^+}\right), \tag{3.11}$$

where $D_P F(\omega(P))$ denotes the coefficient of the linear term in the expansion of a function $F$ in the local parameter in the vicinity of $P$.
In the Newtonian limit this leads to

$$M = \frac{4\Omega^2}{3\pi}, \tag{3.12}$$

the value of the Maclaurin disk, and

$$J = \frac{8\gamma\Omega^3}{15\pi}. \tag{3.13}$$

In the ultrarelativistic limit of the one component disk, $\vartheta_4(u') = 0$, both the mass and the angular momentum diverge. In this limit the dimensionless quotient $M^2/J$ remains bounded and goes to 1, the value of the extreme Kerr metric.

We plot the dimensionless quantity $M^2/J$ in Fig. 7. As a function of $\varepsilon$ it varies monotonically between the Newtonian value

$$\frac{M^2}{J} = \frac{10\Omega}{3\pi\gamma} \tag{3.14}$$

and the value in the ultrarelativistic case which is always bigger than 1 for $\gamma < 1$. For fixed $\varepsilon$ it increases monotonically with $\gamma$.

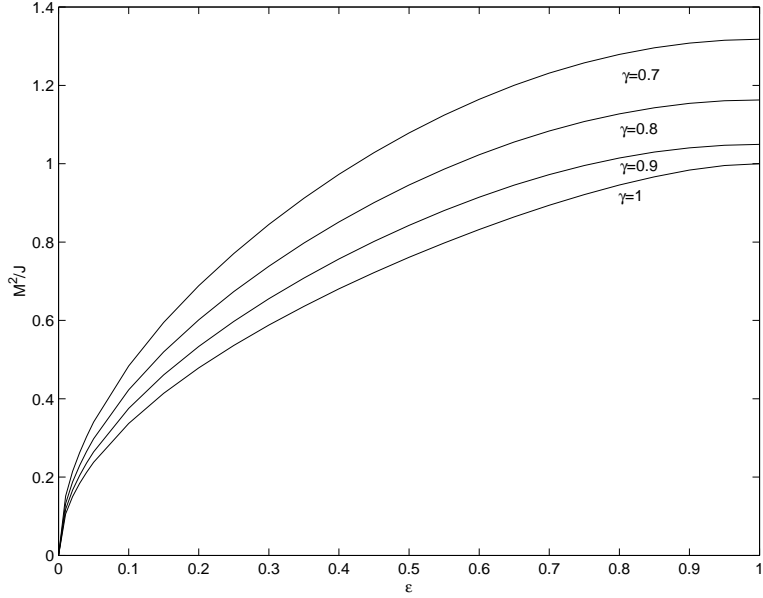

Figure 7: The dimensionless quantity $M^2/J$ in dependence of $\varepsilon$ for several values of $\gamma$.

## 3.3 Energy-momentum tensor

The energy-momentum tensor of the disk is given by (2.5) which has to be considered as an algebraic definition of the tensor components. Since the vectors $u_\pm$ are not normalized, the quantities $\sigma_\pm$ have no direct physical significance. The energy-momentum tensor was chosen in a way to interpolate continuously between the static case and the one-component case with constant angular velocity. An energy-momentum tensor $S^{\mu\nu}$ with three independent components can always be written as

$$S^{\mu\nu} = \sigma_p^* v^\mu v^\nu + p_p^* w^\mu w^\nu, \tag{3.15}$$

where $v$ and $w$ are the unit timelike respectively spacelike vectors $(v^\mu) = N_1(1, 0, \omega_\phi)$ and where $(w^\mu) = N_2(\kappa, 0, 1)$. This corresponds to the introduction of observers (called $\phi$-isotropic observers (FIOs) in [12]) for which the energy-momentum tensor is diagonal. The condition $w_\mu v^\mu = 0$ determines $\kappa$ in terms of $\omega_\phi$ and the metric,

$$\kappa = -\frac{g_{03} + \omega_\phi g_{33}}{g_{00} + \omega_\phi g_{03}}. \tag{3.16}$$

If we introduce the four-velocities $\tilde{u}_\pm = N_\pm u_\pm$, the quantities $\sigma_\pm N_\pm^2$ are proper densities in the sense of [2]. The quantity $\sigma$ which appears in the Einstein equations (see I) is related to $\tilde{\sigma} = \sigma_+ + \sigma_-$ via $\sigma = e^{k-U}\tilde{\sigma}$. In I it was shown that $\sigma$ is given by

$$\sigma = \frac{b_\rho}{8\pi\rho\Omega^2(a-a_0)e^{2U}}. \tag{3.17}$$

It vanishes for $\rho \to 1$ with infinite slope: in the non-static case it was shown in II (Corollary 4.1) that $b_\rho$ is always proportional to $\sqrt{1-\rho^2}$ while in the static case one gets

$$\sigma = \frac{1}{4\pi^2\Omega\left(\frac{\delta}{4}-1+\rho^2\right)} \arctan\sqrt{\frac{1-\rho^2}{\frac{\delta}{4}-1+\rho^2}}. \tag{3.18}$$

Since $b = b_0 + O(\rho^2)$ in the vicinity of the origin for $\varepsilon \neq 1$, the density is regular in the whole disk for $\varepsilon < 1$ and $\gamma \neq 0$. This is however not true in the ultrarelativistic limit of the static disks which we will discuss in more detail in the following section.

The FIOs can interpret the matter in the disk as having a purely azimuthal pressure or as a disk of two counter-rotating dust streams if $p_p^*/\sigma_p^* < 1$. One can show numerically that $p_p^*/\sigma_p^*$ is a monotonically decreasing non-negative function of $\gamma$ which vanishes identically only for $\gamma = 1$. Thus, it is maximal in the static case as expected. There we have

$$1 - \frac{p_p^*}{\sigma_p^*} = 1 - \Omega^2\rho^2 e^{-4U} = e^{2(U_0-U)} \geq 0. \tag{3.19}$$

The last equation follows from (2.25).

The only case where $p_p^* = \sigma_p^*$ is the ultrarelativistic limit of the static disks. In this case the matter rotates with the velocity of light while in all other cases, the velocity $\sqrt{p_p^*/\sigma_p^*}$ is smaller than 1. Thus the energy-momentum tensor can be written in the form

$$S^{\mu\nu} = \frac{1}{2}\sigma_p^*(U_+^\mu U_+^\nu + U_-^\mu U_-^\nu) \tag{3.20}$$

where $(U_\pm^\mu) = U^*(v^\mu \pm \sqrt{p_p^*/\sigma_p^*}\, w^\mu)$ are unit timelike vectors. This is the sum of two energy-momentum tensors for dust. Furthermore it can be shown that the vectors $U_\pm$ are geodesic vectors with respect to the inner geometry of the disk: this is a consequence of the equation $S^{\mu\nu}_{;\nu} = 0$ together with the fact that $U_\pm$ is a linear combination of the Killing vectors. Consequently the FIOs can interpret the matter in the disk as two streams of dust with proper energy density $\sigma_p^*/2$ which are counter-rotating with the same angular velocity $\Omega_c := (N_2/N_1)\sqrt{p_p^*/\sigma_p^*}$. This is the interpretation we will refer to in our discussion.

Except for the static case $\gamma = 0$ the FIOs are not at rest with respect to the locally non-rotating frames which rotate with angular velocity

$$\omega_l := -\frac{g_{03}}{g_{33}} \tag{3.21}$$

with respect to the inertial frame at infinity. Therefore, the quantities we will discuss in the following are the angular velocities $\omega_l$, $\omega_\phi$, $\Omega_c$, and the energy density $\sigma^* := e^{(k-U)}\sigma_p^*$.

We discuss the angular velocities in units of $\Omega$ which has no invariant meaning but which provides a natural scale for the angular velocities in the disk. It is constant with respect to $\rho$ but depends on the parameters $\varepsilon$ and $\gamma$. In the Newtonian limit it is small since $U_0 = -\Omega^2$. Thus independently of $\gamma$, the angular velocity $\Omega$ behaves as $\sqrt{\varepsilon}$ for $\varepsilon \approx 0$. The fact that the ultrarelativistic limit for the one-component disk is reached for a finite value of $\lambda$ implies via (2.6) that $\Omega$ must vanish in this limit. This behavior will be discussed in more detail in section 4. Thus, as $\varepsilon$ varies between 0 and 1, for $\gamma = 1$, $\Omega$ starts near zero in the Newtonian regime, reaches a maximum smaller than 1 and then goes to zero. For $0 < \gamma < 1$, it reaches a maximum, too, but then it does not go to zero in the ultrarelativistic limit. In the static case ($\gamma = 0$) one has

$$\Omega(\varepsilon, 0) = \frac{1}{2}\sqrt{1 - (1-\varepsilon)^4}, \tag{3.22}$$

which grows monotonically from zero to $1/2$ in the ultrarelativistic limit. We plot $\Omega$ as function of $\varepsilon$ for several values of $\gamma$ in Fig. 8.

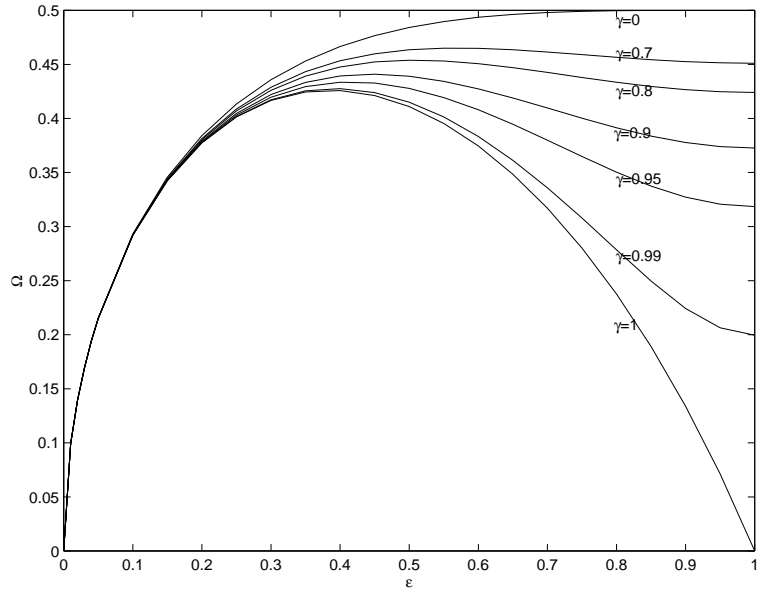

Figure 8: Angular velocity $\Omega$ in dependence of $\varepsilon$ for several values of $\gamma$.

The angular velocity $\omega_l$ of the locally non-rotating observers is a measure for the frame dragging due to the rotating disk. We depict $\omega_l$ in dependence of $\rho$ at the disk for $\gamma = 0.7$ and several values of $\varepsilon$ in Fig. 9. There is obviously no frame dragging in the Newtonian case, $\omega_l$ is of order $\Omega^3$ for small $\Omega$. The angular velocity $\omega_l$ increases monotonically with $\varepsilon$ for fixed $\rho$ and $\gamma$. However the curves for $\varepsilon \geq 0.85$ are so close to the curve with $\varepsilon = 0.85$ that we omitted them in Fig. 9. Since the density (see below) is peeked at the center of the disk for $\varepsilon \to 1$, the frame dragging increases strongly near the center. In Fig. 10 we plot $\omega_l$ at the disk for $\varepsilon = 0.8$ for several values of $\gamma$. In the static case it is identical to zero. The frame dragging increases monotonically with $\gamma$ for fixed $\rho$ and $\varepsilon$ since more counter-rotating matter makes the spacetime more static. Since the central density decreases with $\gamma$ for fixed $\varepsilon$, the frame dragging at the center is for $\gamma < 1$ closer to the one-component case than at the rim of the disk. The angular velocity $\omega_l$ is always smaller than $\Omega$ for $\gamma < 1$. In the ultrarelativistic limit for $\gamma = 1$ the ratio $\omega_l/\Omega$ becomes identical to 1 in the disk.

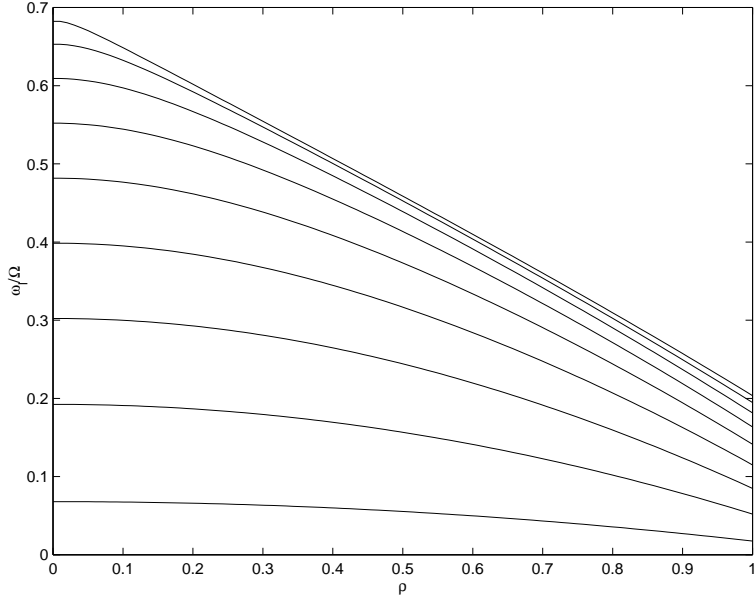

Figure 9: Angular velocity $\omega_l$ for $\gamma = 0.7$ and $\varepsilon = 0.05, 0.15, ..., 0.85$.

In terms of the components of the energy-momentum tensor, the angular velocity $\omega_\phi$ reads

$$\omega_\phi = \frac{1}{2S_3^0}\left(S_3^3 - S_0^0 - \sqrt{(S_3^3 - S_0^0)^2 + 4S_0^3 S_3^0}\right). \tag{3.23}$$

For fixed $\rho$ and $\varepsilon$, the angular velocity $\omega_\phi$ is monotonically increasing in $\gamma$ from zero in the static case to $\Omega$ in the one-component limit. For $\rho = 0$ it is identical to $\gamma\Omega$ which is also the value in the Newtonian limit. The ratio $\omega_\phi/\Omega$ is depicted in dependence of $\rho$ for $\gamma = 0.7$ for several values of $\varepsilon$ in Fig. 11.

The angular velocity of the dust streams $\Omega_c$ with respect to the FIOs follows from

$$\Omega_c = \sqrt{\frac{\omega_\phi^2 - 2\omega_\phi\gamma\Omega + \Omega^2}{1 - 2\kappa\gamma\Omega + \Omega^2\kappa^2}}. \tag{3.24}$$

For fixed $\rho$ and $\varepsilon$ the angular velocity $\Omega_c$ increases monotonically in $\gamma$ from 0 in the one-component case to 1 in the static case. In the former case the observer follows the dust and can interpret the dust which is at rest in his coordinate system as 'two' non-rotating dust components. For $\rho = 0$ the function $\Omega_c$ is identical to $\Omega\sqrt{1-\gamma^2}$ which is also the value in the Newtonian limit. We plot $\Omega_c$ in dependence of $\rho$ for $\gamma = 0.7$ for several values of $\varepsilon$ in Fig. 12.

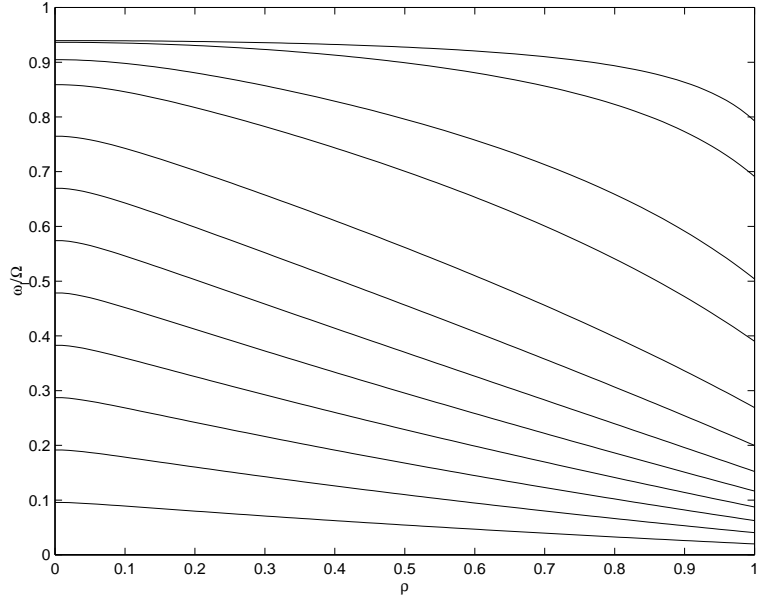

Figure 10: Angular velocity $\omega_l$ for $\varepsilon = 0.8$ and $\gamma = 0.1, 0.2, ..., 0.9, 0.95, 0.99, 1$.

The proper density $\sigma_p^*$ for a FIO is given by

$$\sigma_p^* = \frac{\tilde{\sigma}}{1 - \kappa \omega_\phi} \frac{\rho^2}{\kappa g_{03} + g_{33}} (1 - 2\kappa\gamma\Omega + \kappa^2\Omega^2). \tag{3.25}$$

The density is finite except in the ultrarelativistic limit of the static disks. In the Newtonian limit, the density reads

$$\sigma^* = \tilde{\sigma}(1 + \Omega^2((1 - \gamma^2)\rho^2 - 2)) = \frac{2\Omega^2}{\pi^2}\sqrt{1 - \rho^2} \tag{3.26}$$

the value for the Maclaurin disk. The dependence of $\sigma^*$ on $\rho$ is shown for $\gamma = 0.7$ for several values of $\varepsilon$ in Fig. 13. With increasing $\varepsilon$, the central density grows and the matter is more and more concentrated at the center of the disk. For $\varepsilon = 0.8$ the density is plotted for several values of $\gamma$ in Fig. 14. With increasing $\gamma$, the central density increases.

In [2] and [1] the observer dependent 'rest mass density' $\sigma_{0,\pm}$ of the dust streams was defined as $\sigma_{0,\pm} = \sigma^*/2U_\pm^0$ which leads to the total rest mass density $\sigma_0$ in the asymptotically fixed frame

$$\sigma_0 = \sigma^* \frac{N_1}{U^*}. \tag{3.27}$$

The total rest mass of the disk $M_0$ is then the integral

$$M_0 = 2\pi \int_0^1 \sigma_0 \rho d\rho. \tag{3.28}$$

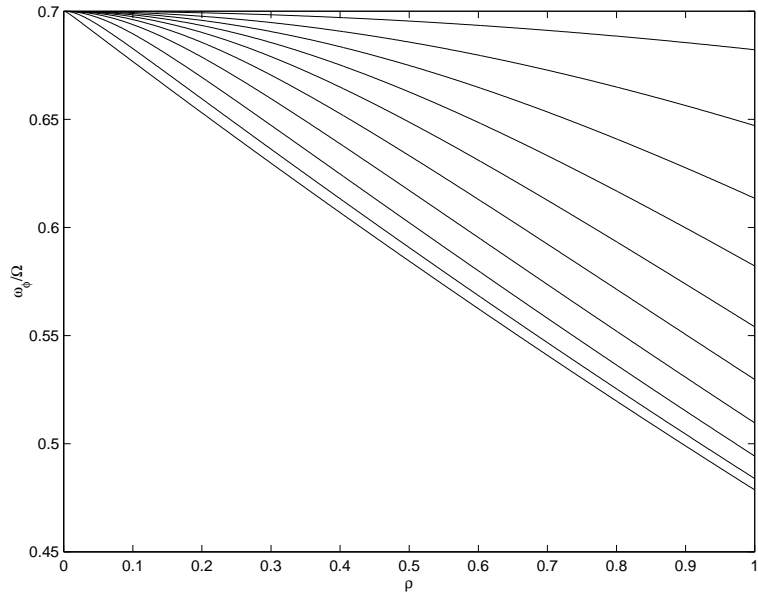

Figure 11: Angular velocity $\omega_\phi$ for $\gamma = 0.7$ and from top to bottom $\varepsilon = 0.05, 0.15, \ldots, 0.95$.

The binding energy of the disk is defined in [2] and [1] as the difference between the total rest mass and the ADM-mass, $E_b = M_0 - M$. We plot $E_b/M_0$ as a function of $\varepsilon$ for several values of $\gamma$ in Fig. 15. In the Newtonian limit, the binding energy is independent of $\gamma$,

$$E_b/M = \frac{1}{5}\Omega^2. \tag{3.29}$$

In the case $\gamma = 1$, the binding energy increases monotonically up to a value of $E_b/M_0 \approx 0.37$ in the ultrarelativistic limit. For $\gamma < 1$ it reaches a maximum for a finite value of $\varepsilon$ and can become even negative. In the static limit $E_b/M_0$ diverges to $-\infty$ in the ultrarelativistic limit since the rest mass of the disk goes to zero. We plot $E_b/M_0$ as function of $\varepsilon$ for several values of $\gamma$ in Fig. 15.

The ADM-mass can also be calculated in standard manner [18] at the disk, in our case

$$M = 2\pi \int_0^1 (S_3^3 - S_0^0) e^{k-U} \rho d\rho. \tag{3.30}$$

Similarly, one gets for the angular momentum

$$J = 2\pi \int_0^1 S_3^0 e^{k-U} \rho d\rho. \tag{3.31}$$

The above formulas can be used to check the numerics since they must reproduce the results of (3.10) and (3.11).

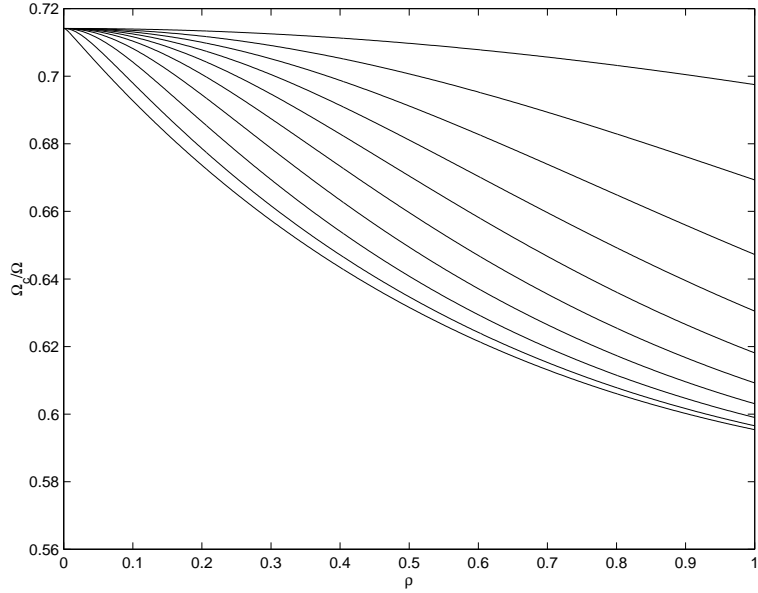

Figure 12: Angular velocity $\Omega_c$ for $\gamma = 0.7$ and and from top to bottom $\varepsilon = 0.05, 0.15, \ldots, 0.95$.

## 3.4 Ergospheres

In strongly relativistic situations it is possible that the asymptotically timelike Killing vector $\partial_t$ becomes null or even spacelike. The vanishing of $e^{2U}$ defines an ergosphere (although it does not have the topology of a sphere here) i.e. the boundary of a region of spacetime where there can be no static observer with respect to infinity.

The surface plot of the metric function $e^{2U}$ in Fig. 2 shows the typical behavior of these functions: they are completely smooth in the exterior of the disk while the normal derivatives are discontinuous at the disk. The function does not assume a local extremum in the exterior of the disk and goes to 1 at infinity, $e^{2U} = 1 - 2M/|z| + \ldots$. Since the ADM-mass is always positive in the physical range of the parameters (see section 4.4), the real part of the Ernst potential is always less than 1. At the disk, however, the function may have a global minimum.

In the Newtonian regime, the so-called gravito-magnetic effects such as ergospheres do not play a role. When the parameter $\varepsilon$ increases from zero to one, the function $e^{2U}$ may vanish at some points in the spacetime. Since it assumes its minimum value at the disk, this means that an ergosphere necessarily first appears at the disk when the minimum value becomes zero. For larger values of $\varepsilon$ the minimum drops below zero in these cases so that the ergosphere grows for increasing values of $\varepsilon$. In the ultrarelativistic limit $\varepsilon = 1$ it reaches the axis.

To illustrate the dependence of ergospheres on the parameter $\varepsilon$ for fixed $\gamma$, we plot them in Fig. 16 for $\gamma = 1$. The plot shows the $(\rho, \zeta)$-plane with the disk on the $\rho$-axis between zero and one. The potential is regular in the equatorial plane in the exterior of the disk which implies that the equipotential surfaces hit the plane orthogonally there. At the disk, however, the normal derivatives have a jump which leads to a cusp of the equipotential contours at the disk. The ergosphere grows with $\varepsilon$ and includes the whole spacetime in the ultrarelativistic limit which will be discussed in the next section.

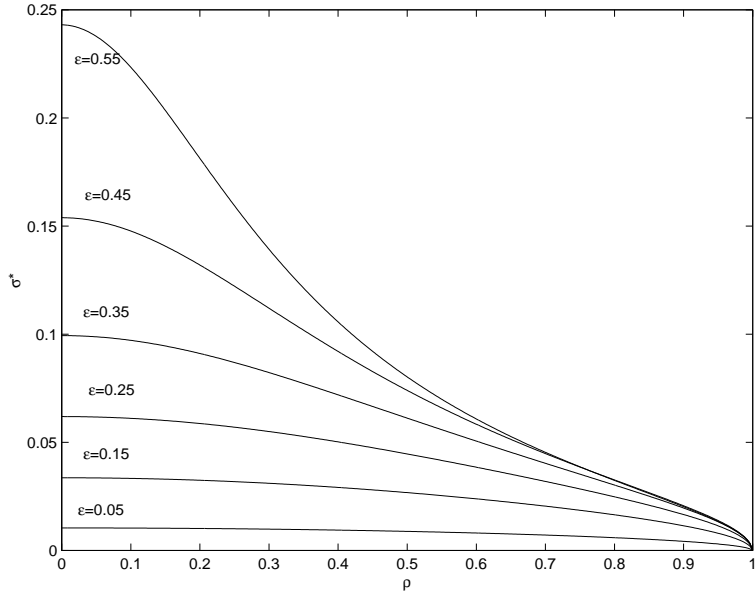

Figure 13: Energy density $\sigma^*$ for $\gamma = 0.7$ and several values of $\varepsilon$.

Qualitatively, one would expect that counter-rotation makes a solution more static, i.e. that effects like ergospheres are suppressed. Thus in situations with the same central redshift but different $\gamma$, the ergoregion will always be smaller in the case of more counter-rotation if there is an ergoregion at all. In Fig. 17 we show the ergospheres for $\varepsilon = 0.95$ and several values of $\gamma$. It follows from (2.26) that the ergosphere goes through the rim of the disk if

$$\delta = 1 - \frac{2}{\lambda}. \tag{3.32}$$

This means that for disks with $\delta > 1$ possible ergoregions are confined to values of $\rho < 1$. One finds numerically that smaller values of $\gamma$ i.e. more counter-rotating matter imply that the ergoregion forms at bigger values of $\varepsilon$ i.e. in stronger relativistic situations if it is to appear at all. The ergoregions are also formed closer to the axis. In the static case there is obviously no ergosphere. The function $e^{2U}$ only vanishes in the ultrarelativistic limit at the center of the disk. There are no ergoregions for values of $\gamma < \gamma_c = 0.707\ldots$.

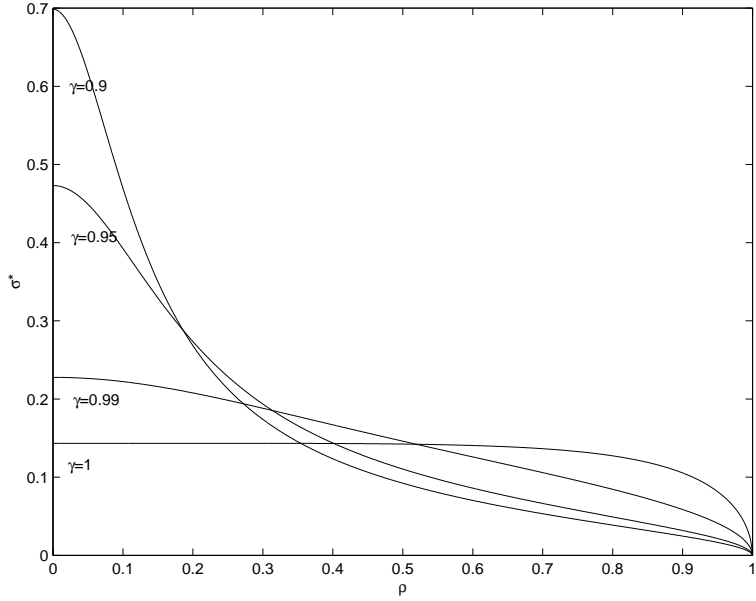

Figure 14: Energy density $\sigma^*$ for $\varepsilon = 0.8$ and several values of $\gamma$.

# 4 Ultrarelativistic limit

## 4.1 Ultrarelativistic limit of the static disks

The main features of the ultrarelativistic limit can already be found in [1]. The potential $e^{2U}$ in the disk and its normal derivative there have the form

$$e^{2U} = \frac{\rho}{2} \quad , \left(e^{2U}\right)_\zeta = \frac{1}{\pi}\arctan\sqrt{\frac{1-\rho^2}{\rho^2}}, \qquad (4.1)$$

whereas the metric function $k$ is of order $\rho^2$ for small $k$. The behavior of the metric functions can be obtained from (3.8) and (2.4). The angular velocity in the disk is $\Omega = 1/2$. The matter in the disk moves with the velocity of light since the four-velocity becomes null in the whole disk. The energy-density $\sigma$ (3.18) diverges at the center as $1/\rho^2$, the density $\sigma^* = -g_{00}\sigma$ diverges as $1/\rho$. The ADM-mass is however finite, $M = 1/(4\pi)$. Since the matter moves with the velocity of light, the rest mass of the disk must vanish. Thus the gravitational binding energy is negative. The linear proper radius

$$\rho_p := \int_0^\rho e^{k-U} d\rho' \qquad (4.2)$$

is finite in the disk since the integrand behaves near the center (see II, Corollary 4.1) as $1/\sqrt{\rho}$ and is finite in the rest of the disk. The proper circumferential radius in the disk,

$$\rho_c = \sqrt{g_{33}(\rho)} = \sqrt{2\rho}, \qquad (4.3)$$

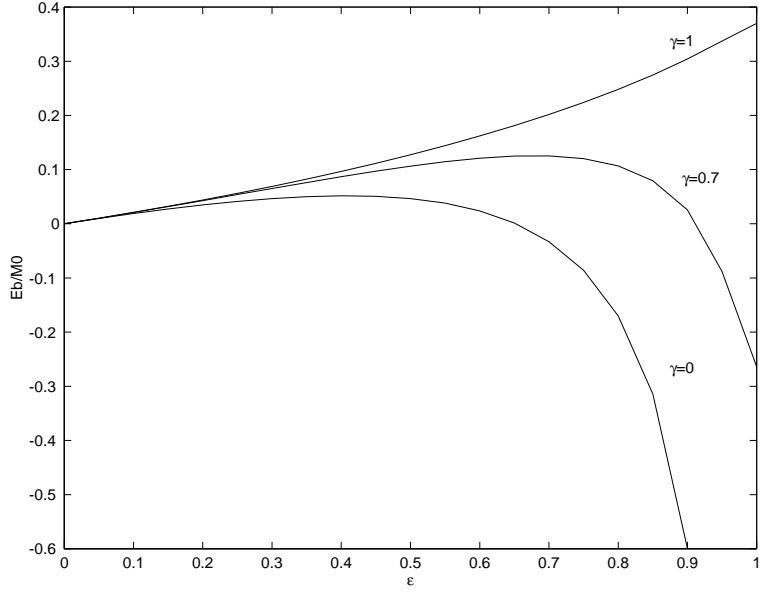

Figure 15: Binding energy of the disks in dependence of $\varepsilon$ for several values of $\gamma$.

is also finite. Thus the ultrarelativistic limit of the static disks with uniform rotation is a disk of finite radius with diverging central redshift and diverging central density but finite mass. The matter in the disk consists of particles with zero rest mass which move with the velocity of light.

## 4.2 Ultrarelativistic limit for $0 < \gamma < 1$

The ultrarelativistic limit of stationary counter-rotating disks bears similarities with the static case in the sense that the axis remains regular: the constants $a_0$ and $C$ in (2.14) and (2.16) which are 0 and 1 respectively in the static case remain finite here since they can only diverge if $\vartheta_4(u') = 0$ which can happen only for $\gamma = 1$. The integrals in the respective exponents of (2.14) and (2.16) are always finite though $\ln G(\tau)$ has a term $\ln \tau$ in the limit $\lambda \to \infty$ as can be easily seen. Thus the axis remains elementary flat in the case $\gamma < 1$ even in the ultrarelativistic limit. Since $a_0 = -\gamma/\Omega$ is non-zero for $0 < \gamma < 1$, the angular velocity $\Omega$ remains finite in the limit, too, as can be seen in Fig. 8.

In II (Corollary 4.1) it was shown that the potential $e^{2U}$ is linear in $\rho$ near the origin unless $\gamma = \gamma_c$ (which is just defined by this condition) where it is quadratic in $\rho$. For $\gamma > \gamma_c$ there are ergospheres in the spacetime, for $\gamma < \gamma_c$ the potential $e^{2U}$ is positive in the whole spacetime. We plot $e^{2U}$ at the disk for several values of $\gamma$ in the ultrarelativistic limit in Fig. 18. We note that the metric function $ae^{2U}$ in the disk is also linear in $\rho$ in the vicinity of the origin if $e^{2U}$ is. For $\gamma \to 0$, the metric function $e^{2U}$ in the disk approaches $\rho/2$. For $\gamma \to 1$ the limiting function is also linear in $\rho$ in the whole disk. One has to note that the limits $\gamma \to 1$ and $\varepsilon \to 1$ do not commute. The ultrarelativistic limit of the case $\gamma = 1$ is discussed section 4.3. The limit $\gamma \to 1$ of the ultrarelativistic solutions for $\gamma < 1$ are always obtained for $\lambda \to \infty$. If one goes with $\gamma \to 1$ ($\delta \to 0$) in this cases, the limiting function is one of the 'overextreme' solutions which are discussed in section 4.4.

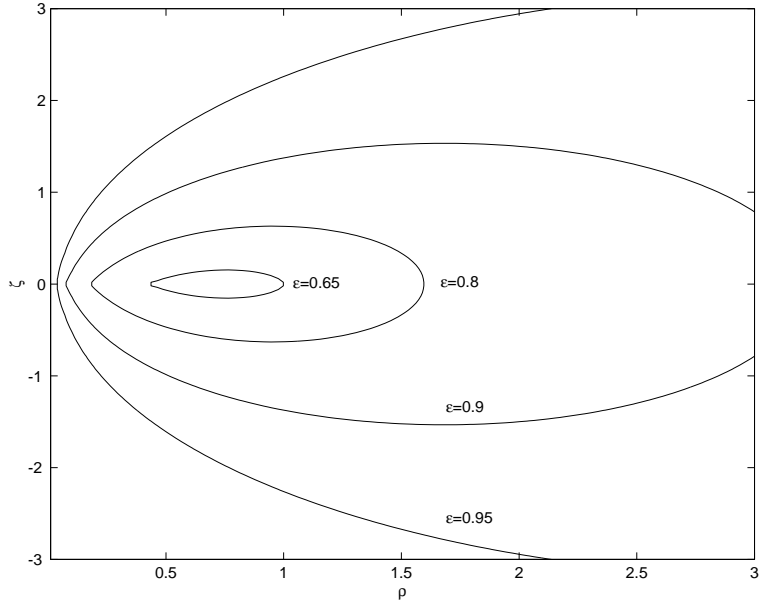

Figure 16: Ergospheres for $\gamma = 1$ and several values of $\varepsilon$.

In contrast to the static case, the energy density $\sigma^*$ is finite even in the ultrarelativistic limit. The proper linear radius (4.2) and the proper circumferential radius (4.3) are both finite in the disk. The velocity of the counter-rotating streams in the disk $\sqrt{p_p^*/\sigma_p^*}$ is less than 1, i.e. the velocity of light in the limit $\varepsilon = 1$ for $0 < \gamma < 1$.

## 4.3 Ultrarelativistic limit of the one-component disks

The ultrarelativistic limit of the case $\gamma = 1$ is different from the previously discussed cases since it is reached for $\vartheta_4(u') = 0$. This implies with (2.14) and (2.16) that both constants $a_0$ and $C$ diverge as $\varepsilon \to 1$. These constants do not have a direct physical importance. The fact that they diverge merely indicates that the axis cannot remain elementary flat in the ultrarelativistic limit. A consequence of the diverging constant $a_0$ is that the angular velocity $\Omega$, which is the coordinate angular velocity in the disk as measured from infinity, vanishes. A diverging constant $C$ implies that all linear proper distances (4.2) diverge. The function $e^{2(k-U)+2U_0}$ is however bounded.

The axis is in fact singular in the sense that the metric function $e^{2U}$ vanishes there identically which can be seen from (2.18). The Ernst potential is identical to $-\mathrm{i}$ on the axis for $\zeta > 0$. In the limit $\varepsilon \to 1$, the ergosphere becomes bigger and bigger. When it finally hits the axis for $\varepsilon = 1$, the whole axis and infinity form the ergosphere and the function $e^{2U}$ is negative in the remainder of the spacetime. We plot the potential in Fig. 19. The fact that $e^{2U}$ vanishes on the whole axis implies moreover that all multipole moments diverge. The dimensionless quotient $M^2/J$ remains however finite and tends to 1, the value of the extreme Kerr metric (see section

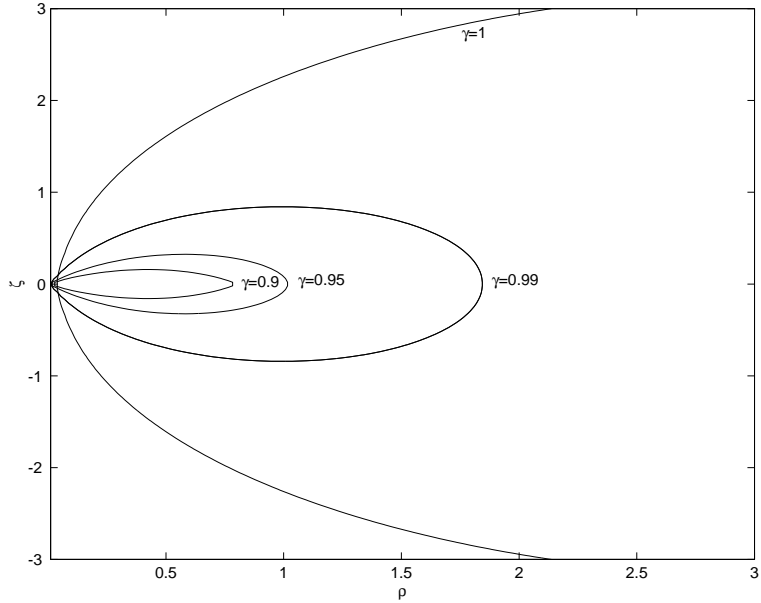

Figure 17: Ergospheres for $\varepsilon = 0.95$ and several values of $\gamma$.

3.4).

The vanishing of $\Omega = \Omega\rho_0$ in the limit $\varepsilon = 1$ indicates that either the angular velocity or the radius of the disk go to zero in this case. Bardeen and Wagoner [2] argued that the spacetime can be interpreted in the limit $\varepsilon \to 1$ and $\rho_0 \to 0$ as the extreme Kerr metric in the exterior of the disk. In [10] it was shown that such a limit (diverging multipoles, singular axis,...) can occur in general hyperelliptic solutions and can always be interpreted as an extreme Kerr spacetime. For an algebraic treatment of the ultrarelativistic limit of the Bardeen-Wagoner disk see [20]. In the ultrarelativistic limit of the above disks for $\gamma = 1$, the spacetime becomes an extreme Kerr spacetime with $m = \frac{1}{2\Omega}$. The physical interpretation of this fact as already given in [2] is that the disks become more and more redshifted for increasing $\varepsilon$. Its radius shrinks and the disk finally vanishes behind the horizon of the extreme Kerr metric which forms in the ultrarelativistic limit.

## 4.4 Over-extreme Region

Since the ultrarelativistic limit of the one-component disks is reached for a finite value $\lambda_c$ of $\lambda$, the question arises what the solution (2.8) describes for $\lambda > \lambda_c$, the smallest value of $\lambda$ where $\varepsilon = 1$. In I it was shown that the boundary conditions at the disk are still satisfied. Moreover the relations between the metric functions at the disk ensure that the functions are bounded at the disk (they have at most a jump discontinuity there). The proof for global regularity given in I does not hold in the 'over-extreme' region $\lambda > \lambda_c$. It indicates that a singularity in the equatorial plane is probable which in fact can be verified numerically. A typical plot is presented in Fig. 20. In the ultrarelativistic limit, the ergosphere stretches to infinity, in the over-extreme region with $\varepsilon < 1$ it is confined to a finite region of spacetime. The singularity in the equatorial plane is of the form $1/(\rho - \rho_s)$ at $\rho_s$ since the elliptic theta functions in the equatorial plane have zeros of first order. In [10] it was shown that the singularity leads to a negative ADM-mass for certain $\lambda > \lambda_c$. The spacetime is thus physically unacceptable. This is a striking example that it is not sufficient to solve a boundary value problem locally at the disk within the class of solutions [4], but that one has to find in addition the range of the physical parameters where the solution is globally regular outside the disk.

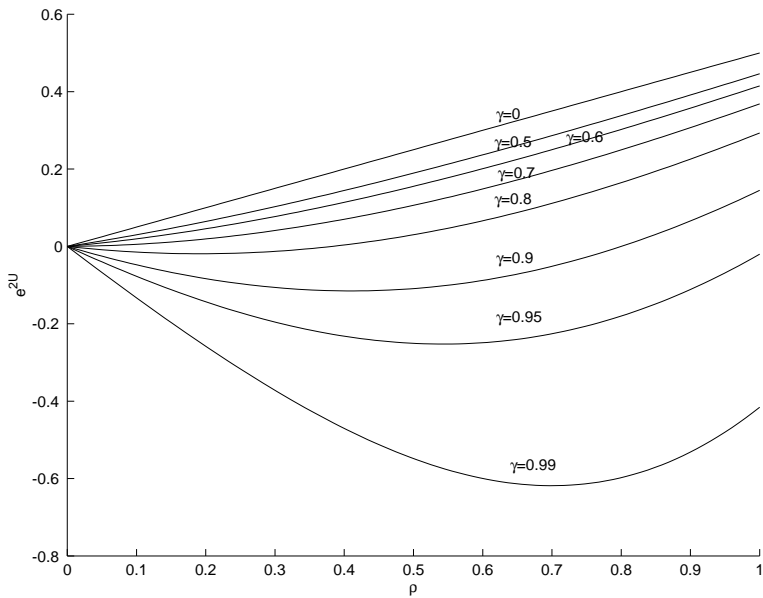

Figure 18: Metric function $e^{2U}$ at the disk for several values of $\gamma$.

# 5 Conclusion

In this paper we have discussed a class of solutions to the Ernst equation which can be interpreted as counter-rotating disks of dust. The solutions are given on a Riemann surface of genus 2. We presented the numerical evaluation of the explicit formulas for the mass and angular momentum, the energy-density, angular velocities in the disk in terms of theta functions. Most of these relations hold for general solutions on Riemann surfaces of genus 2. A generalization to arbitrary finite genus is straight forward in most cases. The discussion here is intended to provide an example on how to extract physical information out of the solutions of the form (2.8). Of special interest is the ultrarelativistic limit in which the redshift of photons emerging from the center of the disk diverges. In the case of only one component, the disk shrinks to a point and the exterior of the solution can be interpreted as the extreme Kerr solution. If counter-rotating matter is present, the disk has a always a finite radius even in the ultrarelativistic limit. It would be interesting to study numerically the light cone structure of the spacetime which will be the subject of further research.

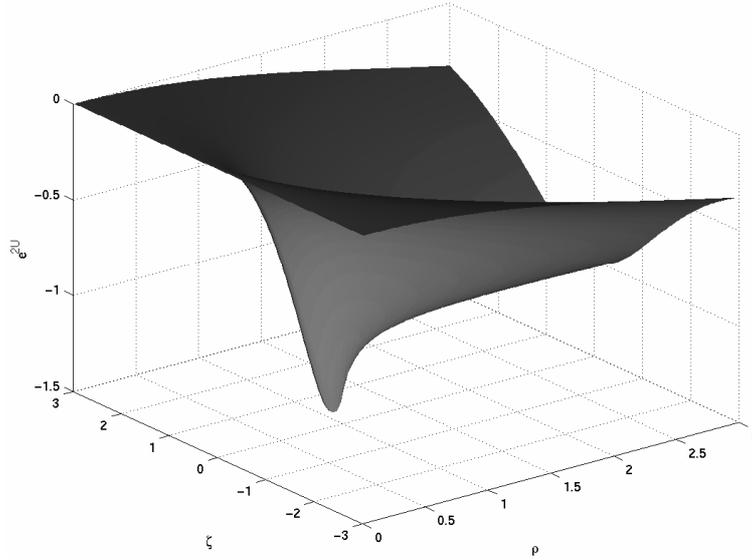

Figure 19: Metric function $e^{2U}$ in the ultrarelativistic limit of $\gamma = 1$.

*Acknowledgment*

We thank A. Bobenko, R. Kerner, D. Korotkin, H. Pfister, O. Richter and U. Schaudt for helpful remarks and hints. One of us (CK) acknowledges support by the Marie-Curie program of the European Community.

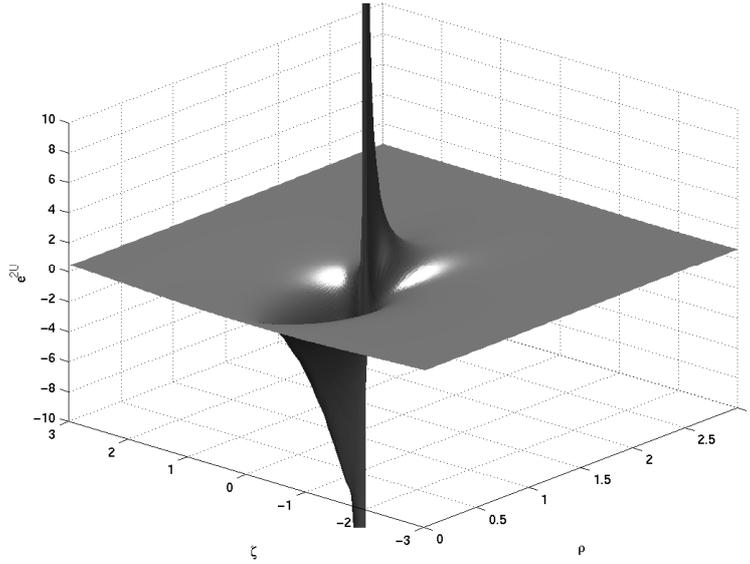

Figure 20: Metric function $e^{2U}$ in the over-extreme region of $\gamma = 1$.